\documentclass[useAMS,usenatbib]{mn2e}
\usepackage{psfig, epsf, epsfig}
\title[Stripping of galactic halo gas]{Ram pressure stripping of halo
gas in disk galaxies:
Implications on  galactic star formation in different environments}
\author[K. Bekki]{Kenji Bekki${}^1$\thanks{E-mail:
bekki@phys.unsw.edu.au}\\
       ${}^1$School of Physics, University of New South Wales,
              Sydney NSW, 2052, Australia}

\begin{document}

\date{Accepted, Received 2005 February 20; in original form }

\pagerange{\pageref{firstpage}--\pageref{lastpage}} \pubyear{2005}

\maketitle

\label{firstpage}

\begin{abstract}

We numerically investigate evolution of gaseous halos around disk
galaxies in different environments ranging from small groups to
rich clusters in order to understand galaxy evolution in these
environments.  Our simulations self-consistently incorporate 
effects of ram pressure of intergalactic medium (IGM) on disk and
halo gas of galaxies and hydrodynamical interaction between disk and halo gas
so that mass fractions of halos gas stripped by
ram pressure of IGM ($F_{\rm strip}$)
can be better estimated.
We mainly investigate how $F_{\rm strip}$ 
depends on
total  masses of their host environments
($M_{\rm host}$), galactic masses ($M_{\rm gal}$),
densities and temperature of IGM ($T_{\rm IGM}$ and ${\rho}_{\rm IGM}$,
respectively), relative velocities between IGM and galaxies ($V_{\rm
r}$), and physical properties of disks (e.g., gas mass fraction).
We find that typically $60-80$\% of halo gas can be efficiently
stripped from Milky Way-type disk galaxies 
by ram pressure in clusters with $M_{\rm host}\sim 10^{14} {\rm M}_{\odot}$.
We also find that $F_{\rm strip}$  depends  on 
$M_{\rm host}$
such that  $F_{\rm strip}$ is higher for larger $M_{\rm host}$.
Furthermore it is found that
$F_{\rm strip}$ can be  higher in
disk galaxies with smaller $M_{\rm gal}$
for a given environment.
Our simulations demonstrate  that the presence of disk gas can suppress 
ram pressure stripping of halo gas  owing to hydrodynamical interaction
between halo and disk gas.
Ram pressure stripping of halo gas is found to be efficient
(i.e., $F_{\rm strip}>0.5$)   even in small and/or
compact groups, if ${\rho}_{\rm IGM}$  $ \sim 10^5 {\rm
M}_{\odot}$ kpc$^{-3}$  and $V_{\rm r}\sim 400$ km s$^{-1}$.
Based on the derived
radial distributions of remaining halo gas after ram pressure
stripping,
we propose that truncation of star formation after halo gas stripping
can occur outside-in in disk galaxies.
We suggest that although gradual truncation of star formation
in disk galaxies
can occur in groups,  it can proceed less rapidly
in comparison with cluster environments. 
We also suggest  
low-mass galaxies are likely to truncate their star formation more
rapidly
owing to more efficient halo gas stripping in groups and clusters.

\end{abstract}

\begin{keywords}
galaxies:halo --
galaxies:structure --
galaxies:kinematics and dynamics 
\end{keywords}

\section{Introduction}

Since Larson et al. (1980) discussed removal of galactic halo gas
from disk galaxies
in terms of transformation from spirals into S0s,
many observational and theoretical works 
investigated  evolution of galactic halo gas 
in variously different aspects of galaxy evolution,
such as maintenance of spiral arms by halo gas infall (e.g., Sellwood
\& Carlberg 1984), formation of passive spiral galaxies
in groups and clusters (e.g., Bekki et al. 2002),
and color evolution of satellite galaxies entering into their host group-
and cluster-scale  
halos (e.g., Font et al. 2008).
One of important suggestions from these previous works is
that stripping of galactic halo gas can cause severe suppression
of galactic global  star formation in different environments
(``strangulation''; Balogh et al. 1999, 2000).
Although the strangulation scenario  can explain a number of recent
observational results of galaxies, such as the presence of
red passive spirals in distant clusters (e.g., Couch et al.  1998,
Dressler et al. 1999,
Poggianti et al. 1999, 2008)
and mean star formation rate dependent on galaxy environments 
(e.g. Balogh et al. 2004),
it is still less clear why and how effective strangulation can occur
in different environments (e.g., groups and clusters).

Previous theoretical and numerical studies
tried to understand how galactic halo
gas responds to environmental effects, such as tidal fields of groups
and clusters (e.g., Bekki et al. 2001) and ram pressure stripping
of IGM (e.g., Balogh et al. 2000; Bekki et al. 2002; 
Hester 2006; McCarthy et al.
2008). Although these works discussed quantitatively
how much fraction of halo gas can be removed from galaxy-scale
halos by ram pressure of IGM in their host environments, 
their models are not so sophisticated at some points.
For example,  previous numerical models (e.g., Bekki et al. 2002)
do not adopt 
realistic galaxy models with stellar and gaseous disks 
and ignores hydrodynamical interaction between halo and disk
gas.  More sophisticated numerical models are required to
discuss more qualitatively how much fraction of halo gas
can be removed by ram pressure stripping and whether and how
hydrodynamical interaction between halo and disk gas can influence
ram pressure stripping of halo gas in disk galaxies.

\begin{table*}
\centering
\begin{minipage}{185mm}
\caption{The ranges of model parameters.}
\begin{tabular}{ccccccc}
Parameters &
{$M_{\rm gal}$ ($\times M_{\rm MW}$)
\footnote{The total mass of a galaxy in units of
$M_{\rm MW}$ ($=1.1 \times 10^{12} {\rm M}_{\odot}$).}}  &
{$V_{\rm r}$ (km s$^{-1}$)
\footnote{The relative velocity of the hot IGM with respect to
the galaxy.}}  &
{${\rho}_{\rm IGM}$ ($\times 10^5 {\rm M}_{\odot}$ kpc$^{-3}$)
\footnote{The initial mass density of the hot
IGM in the cube.}}  &
{$T_{\rm IGM}$ ($\times 10^7$ K)
\footnote{The initial temperature  of the hot
IGM in the cube.}}  &
{$f_{\rm g}$ 
\footnote{The initial gas mass fraction
in a galactic disk.}}    &
{$R_{\rm halo}$ ($\times R_{\rm d}$) 
\footnote{The initial radius of a spherical galactic halo gas  
of  a disk galaxy in units of $R_{\rm d}$, where $R_{\rm d}$
is the disk size.}}   \\
Value ranges & 0.05--1.0 & 160--2000 & 0.1--5.0 & 0.1--3.2  & 0--0.1 &
1--6\\
\end{tabular}
\end{minipage}
\end{table*}

The purpose of this paper is thus to investigate time evolution of
galactic halo gas in disk galaxies under moderately strong
ram pressure of IGM in their host environments based on
more sophisticated numerical simulations.
We focus mainly on (i) final mass fractions of halo gas stripped from
disk galaxies ($F_{\rm strip}$),
(ii) radial properties of halo gas, and
(iii) dependences of (i) and (ii) on physical properties
of galaxy environments (e.g., total masses of groups and clusters). 
The present results can be used for interpreting observational results
on spatial distribution of hot  halo gas in galaxies
(e.g., Jeltema et al. 2008),
physical properties of passive spirals in  groups and clusters
(e.g., Poggianti et al. 2008; Bamford et al. 2009), and
origin of possible suppression of star formation in compact groups
(e.g., Rasmussen et al. 2008).

The plan of the paper is as follows: In the next section,
we describe our  numerical models  for ram pressure stripping of
galactic
halo gas. 
In \S 3, we
present the numerical results
mainly on the physical properties of  remaining halo gas after the ram pressure
stripping
for variously different models.
In \S 4, we compare the present results with those from other authors
and discuss implications of the present results.
We summarize our  conclusions in \S 5.
We here do not discuss ram pressure stripping of {\it disk gas}
in galaxies in an extensive manner, because a number of authors
have already discussed the stripping processes in detail
(e.g., Abadi et al. 1999;
Kronberger et al. 2008; 
Marcoline et al. 2003;
Roediger \& Bruggen 2008;
Tonnesen \& Bryan 2008;  Vollmer et al. 2006).

\begin{figure}
\psfig{file=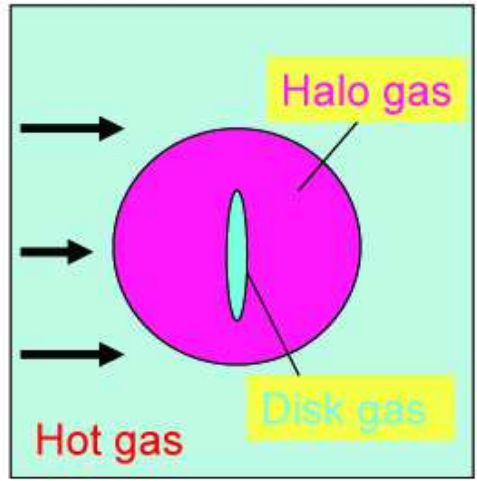,width=7.0cm}
\caption{
A schematic representation of the present numerical model
for ram pressure stripping of galactic halo gas in disk galaxies.
Both spherical gaseous halo and initially thin gaseous disk in 
a disk galaxy can be influenced by hot IGM in groups and clusters
of galaxies in the present model. Both hydrodynamical interaction
between halo and disk gas and that between hot IGM and disk and halo
gas are modeled in a self-consistent manner in the preset study.
}
\label{Figure. 1}
\end{figure}

\begin{figure*}
\psfig{file=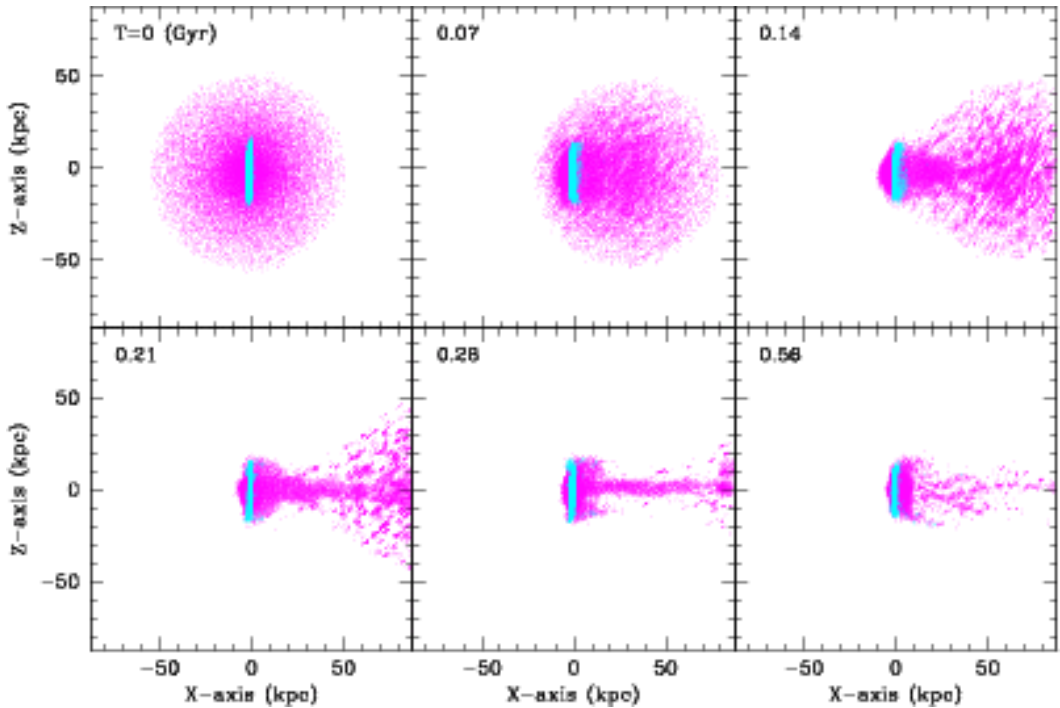,width=14.0cm}
\caption{
Morphological evolution of halo (magenta) and disk (cyan)  gas
projected onto the $x$-$z$ plan
in a disk galaxy for the standard model. The dark matter halo,
stellar disk, bulge, and hot IGM are not shown for clarity.
The time $T$ in units of Gyr is shown in the upper left corner
for each frame.
}
\label{Figure. 2}
\end{figure*}

\section{The model}

\subsection{Disk galaxy}

In order to simulate the time evolution of
galactic halo gas under ram pressure of IGM,
we use the latest version of GRAPE
(GRavity PipE, GRAPE-7) which is the special-purpose
computer for gravitational dynamics (Sugimoto et al. 1990).
We use our original GRAPE-SPH code (Bekki \& Chiba 2006)
which combines
the method of smoothed particle
hydrodynamics (SPH) with GRAPE for calculations of three-dimensional
self-gravitating fluids in astrophysics. The original code used
in our previous studies is here
revised so that both (i) ram pressure effects of hot IGM on galactic
halo gas and (ii) hydrodynamical 
interaction between halo and disk gas can be investigated
in a fully self-consistent manner.
Fig. 1 illustrates initial configurations for  halo and disk gas in a
disk galaxy  and hot IGM surrounding the disk galaxy in
the present numerical study.

Since our numerical methods for modeling dynamical evolution
of  Milky Way-type disk galaxies have already been described by
Bekki \& Shioya (1998) 
and by Bekki \& Peng (2006), we give only a brief
review here.
The total disk mass and the size of a disk of a Milky Way-type disk galaxy
with the total mass of $M_{\rm gal}$
are $M_{\rm d}$ and $R_{\rm d}$, respectively.
Henceforth, all masses and lengths are measured in units
of $M_{\rm d}$ and $R_{\rm d}$, respectively, unless
specified. Velocity and time are measured in units of $v$ = $
(GM_{\rm d}/R_{\rm d})^{1/2}$ and $t_{\rm dyn}$ = $(R_{\rm
d}^{3}/GM_{\rm d})^{1/2}$, respectively, where $G$ is the
gravitational constant and assumed to be 1.0 in the present
study. If we adopt $M_{\rm d}$ = 6.0 $\times$ $10^{10}$ $ \rm
M_{\odot}$ and $R_{\rm d}$ = 17.5 kpc as a fiducial value, then
$v$ = 1.21 $\times$ $10^{2}$ km/s and $t_{\rm dyn}$ = 1.41
$\times$ $10^{8}$ yr, respectively.
The disk is composed of a dark matter halo,
a stellar disk, a stellar bulge, a gaseous disk,
and a gaseous halo.

The mass ratio of the dark matter halo to the stellar disk
in a disk model is
fixed at 16.7 models (i.e., $M_{\rm gal}/M_{\rm d}=17.7$).
We adopt the density distribution of the NFW
halo (Navarro, Frenk \& White 1996) suggested from CDM simulations:
\begin{equation}
{\rho}(r)=\frac{\rho_{0}}{(r/r_{\rm s})(1+r/r_{\rm s})^2},
\end{equation}
where  $r$, $\rho_{0}$, and $r_{\rm s}$ are
the spherical radius,  the characteristic  density of a dark halo,  and the
scale
length of the halo, respectively.
The value of $r_{\rm s}$ (0.6 in our units for $c$ = 10) is chosen such
that
the rotation curve of a disk is reasonably consistent with observations.
The mass fraction  and
the scale length of the stellar bulge represented by the Hernquist
profile
are fixed at 0.17
(i.e., 17 \% of the stellar disk)
and 0.04 (i.e., 20 \%) of the scale length of the stellar disk),
respectively, which are consistent with those of the bulge
model of the Galaxy.

The radial ($R$) and vertical ($Z$) density profiles of the disk are
assumed to be proportional to $\exp (-R/R_{0}) $ with scale
length $R_{0}$ = 0.2 and to ${\rm sech}^2 (Z/Z_{0})$ with scale
length $Z_{0}$ = 0.04 in our units, respectively: both stellar and
gaseous disks follow this exponential distribution.
In addition to the   
rotational velocity caused by the gravitational field of disk,
bulge, and dark halo components, the initial radial and azimuthal
velocity dispersions are assigned to the disc component according to
the epicyclic theory with Toomre's parameter $Q$ = 1.5.  The
vertical velocity dispersion at given radius is set to be 0.5
times as large as the radial velocity dispersion at that point,
as is consistent with the observed trend of the Milky Way (e.g.,
Wielen 1977).

We investigate  models with different $M_{\rm gal}$ and adopt
the Freeman's law (Freeman 1970) to determine $R_0$ 
of a disk galaxy according to its disk mass:
\begin{equation}
R_{\rm 0}=3.5 {(\frac{M_{\rm d}}{6\times 10^{10} {\rm
M}_{\odot}})}^{0.5} {\rm kpc.}
\end{equation}
Structural and kinematical properties of dark matter halos
and stellar disks are assumed to be self-similar between
models with different $M_{\rm gal}$.
The gas mass fraction ($f_{\rm g}$) is assumed
to be a free parameter. 
An isothermal equation of state is used for the gas with
temperatures of $10^4$ K for models with $M_{\rm d}=6 \times
10^{10} {\rm M}_{\odot}$. The initial temperature ($T_{\rm iso}$) of disk
gas
is assumed to be scaled to $T_{\rm iso} \propto {M_{\rm gal}}^{0.5}$.

\subsection{Halo gas and hot IGM}

The
gaseous halo has mass $M_{\rm hg}$
and the same spatial distribution as the dark
matter and is assumed to be initially in hydrostatic equilibrium. The
initial
gaseous temperature of a halo gas particle is therefore determined by
the gas
density, total mass, and gravitational potential at the location of the
particle via Euler's equation for hydrostatic equilibrium (e.g., the
equation
1E-8 in Binney \& Tremaine 1987).
Therefore  gaseous temperature $T_{halo}(r)$ at radius $r$ from the center
of a disk galaxy can be described as:
\begin{equation}
T_{\rm halo}(r)=
\frac{m_{\rm p}}{k_{\rm B}}
\frac{1}{{\rho}_{\rm halo}}
\int_{r}^{\infty}
{\rho}_{\rm halo}(r)
\frac{GM(r)}{r^2}
dr,
\end{equation}
where $m_{\rm p}$ $G$, and $k_{\rm B}$ are the proton mass, 
the gravitational constant, and the Boltzmann constant, respectively,
and $M(r)$ is the total mass within $r$ determined by
the adopted mass distributions of dark matter and baryonic
components in the disk galaxy. 
Radiative cooling is not included in the present study so that
the hydrodynamical equilibrium of halo gas can be obtained
in isolated disk models:if gaseous cooling is included in a disk model,
the halo gas can rapidly collapse to settle down onto the gas disk
so that physical roles of ram pressure stripping in evolution
of halo gas can not properly investigated.
The present idealized models thus help us to grasp some
essential ingredients of the physical roles of hot IGM 
in evolution of galactic halo gas.

Sembach et al. (2003) showed that
the Galaxy gaseous halo is highly extended ($R\sim 70$ kpc)
and low-density (${\rho}_{\rm halo} \approx 10^{-4} - 10^{-5}$ cm$^{-3}$).
It is however observationally unclear how far galactic halo gas 
extends in other galaxies.
We therefore consider that the radius
of the galactic halo ($R_{\rm halo}$) is a free parameter.
Guided by the above observational results by Sembach et al. (2003),
we mainly investigate models with
$R_{\rm halo}=3-6R_{\rm d}$ 
and  ${\rho}_{\rm halo} = 10^{-4}$ cm$^{-3}$.

The disk galaxy is assumed to be embedded in  hot IGM with
the temperature of $T_{\rm IGM}$  and the
densities of ${\rho}_{\rm IGM}$.  In order to avoid huge
particle numbers to represent the entire IGM in group and clusters
of galaxies (e.g., Abadi et al. 1999),
the IGM is represented by SPH particles with  velocities of
$V_{\rm r}$ in a cube with the size of $R_{\rm IGM}$.
The initial velocity of each SPH particle for the IGM
is set to be ($V_{\rm r}$, 0, 0) for all models
(i.e., the particle flows along the $x$-axis
to the positive $x$ direction).
The IGM  has an uniform distribution 
within the cube and $R_{\rm IGM}$ is set to be $10M_{\rm d}$ 
for models with $R_{\rm halo}=3R_{\rm d}$.
We include periodic boundary conditions (at $R_{\rm IGM}$)
for the IGM 
SPH particles leaving the cube.
The parameter values of $T_{\rm IGM}$ and ${\rho}_{\rm IGM}$ are
chosen based on the adopted total masses of galaxy's host  environment
($M_{\rm host}$).

The spin of the disk  galaxy 
is specified by two angles $\theta$ and
$\phi$ (in units of degrees), where
$\theta$ is the angle between
the $z$-axis and the vector of the angular momentum of the disk,
and $\phi$ is the azimuthal angle measured from $x$ axis to
the projection of the angular momentum vector of the disk onto
the $x$-$y$ plane.
In order to show more clearly the ram pressure effects on
galactic halo gas,  we mainly investigate the models with $\theta=90^{\circ}$
and $\phi=0^{\circ}$, in which the 
the entire gas disk can strongly feel ram pressure of IGM. 
We also investigate the models with different $\theta$ and $\phi$
to clarify the roles of halo-disk hydrodynamical interaction
in keeping halo gas within disk galaxies.

The mass resolution
for disk gas particles, halo ones, and IGM
in  luminous disk
models with $M_{\rm d}=6 \times 10^{10}
{\rm M}_{\odot}$ are $3.5 \times 10^5 {\rm M}_{\odot}$,
$5.9 \times 10^4 {\rm M}_{\odot}$
and $5.3 \times 10^5 {\rm M}_{\odot}$, respectively.
The typical smoothing length for disk  gas particles
(the mean smoothing length at $T=0.21$ Gyr)
in the standard model are 680pc (this can be as large as 4.7 kpc
for halo gas particles owing to the stripped halo gas).

\subsection{Parameter study}

Although we have investigated 45 models with different
$M_{\rm gal}$, $f_{\rm g}$, $V_{\rm r}$, ${\rho}_{\rm IGM}$,
and $T_{\rm IGM}$, we mainly show the results of the ``standard''
model with $M_{\rm gal} = 1.1 \times 10^{12} {\rm M}_{\odot}$,
$f_{\rm g}=0.1$,  $V_{\rm r}=500$ km s$^{-1}$,
${\rho}_{\rm IGM}=10^5 {\rm M}_{\odot}$ kpc$^{-3}$,
and  $T_{\rm IGM}=10^7$ K.
This is mainly because the standard model can clearly show
a typical behavior of 
ram pressure stripping of halo gas from disk galaxies.
For a canonical baryonic mass fraction of 0.14 
($= {\rm \Omega}_{\rm b}/{\rm \Omega}_{\rm m}$)  in the universe,
${\rho}_{\rm IGM}$ can be as high as $10^5 {\rm M}_{\odot}$ kpc$^{-3}$
at $r\sim100$ kpc in a cluster with the NFW profile,
the total mass of $10^{14} {\rm M}_{\odot}$,
and the virial velocity ($v_{\rm vir}$) of $500-600$ km s$^{-1}$.
We thus consider that
the adopted values of ${\rho}_{\rm IGM}$,
$V_{\rm r}$, and $T_{\rm IGM}$ are reasonable for
a cluster of galaxies with $M_{\rm host}=10^{14} {\rm M}_{\odot}$.
It should be stressed here that the strength of ram pressure
($P_{\rm ram}$)
in the standard model is much weaker than that required for
ram pressure stripping of disk gas in 
{\it the central regions of clusters}
(e.g., Abadi et al. 1999).
The adopted galaxy mass in the standard model is referred
to as $M_{\rm MW}$ just for convenience.

The total  number of particles used for the disk galaxy
and the IGM in the standard model
are 133699 and 125000, respectively.
The present simulation requires typically $70-160$  CPU hours of
the adopted  GRAPE-7 
systems for each model
depending on model parameters:
we have adopted the above  particle numbers
to run numerous models within a reasonable time scale.
We have conducted a resolution test by using a high-resolution model
with the total particle number of 333805  and found
that there is only $\sim 7$\% difference in $F_{\rm strip}$
between the standard model and the high-resolution one. 
We thus consider that the above particle number is enough
to discuss ram pressure stripping of halo gas around disk galaxies.
Gravitational softening lengths in simulation units ($R_{\rm d}$)
are set to be
fixed at 0.018 for the  galaxy and  0.2 for the IGM.
The range of model parameters investigated in the present study
are shown in the Table 1.

We divide models into four categories: ``Rich cluster'' model
with $M_{\rm host}=10^{15} {\rm M}_{\odot}$, 
``cluster'' one with 
$M_{\rm host}=10^{14} {\rm M}_{\odot}$,
``group'' one with 
$M_{\rm host}=10^{13} {\rm M}_{\odot}$,
and ``small group'' one with 
$M_{\rm host}=10^{12} {\rm M}_{\odot}$.
Considering virial velocities ($v_{\rm vir}$) of dark matter  halos (NFW95)
and the scaling relation between masses and sizes for dark matter halos
(e.g., Padmanabhan 1993),
reasonable values of $T_{\rm IGM}$ and $V_{\rm r}$
are chosen for the above four models with different $M_{\rm host}$.
In the present study,  $V_{\rm r}$ of IGM in
a galaxy  environment with $M_{\rm host}$ is set to be similar to
$v_{\rm vir}$ of  a dark matter halo with $M_{\rm host}$.
$T_{\rm IGM}$ in the small group, group, cluster, and rich-cluster
models are  
$10^6$ K,
$3.2 \times 10^6$ K,
$10^7$ K,
and $3.2 \times 10^7$ K,  respectively, in the present study.
$V_{\rm r}$ in the small group, group, cluster, and rich-cluster
are 
158 km s$^{-1}$, 
281 km s$^{-1}$, 
500 km s$^{-1}$, 
and 889 km s$^{-1}$, respectively.

We mainly investigate the time evolution of gas mass ($M_{\rm g}$) 
for halo and disk components in galaxies and the final mass fraction
of halo gas stripped from  galaxies ($F_{\rm strip}$). We consider 
that if SPH gas particles at the final time step of a simulation
are outside $R_{\rm halo}$, they are regarded as being stripped by
ram pressure of IGM. For most models, halo gas can be rapidly stripped
from galaxies within well less than 0.5 Gyr. Therefore,
we estimate $F_{\rm strip}$ at $T=0.56$ Gyr, where 
the  time $T$ represents the time that has elapsed
since
the simulation starts.
We also estimate the  accretion rate of gas onto the central 1kpc
of a disk galaxy in a simulation by dividing the total gas mass
accumulated in the central 1 kpc at the final time step
by time that has elapsed since the simulation starts (i.e., 0.56 Gyr).
We confirm that
the  accretion rates in different models do not depend so much
on model parameters in the present study. 

\begin{figure}
\psfig{file=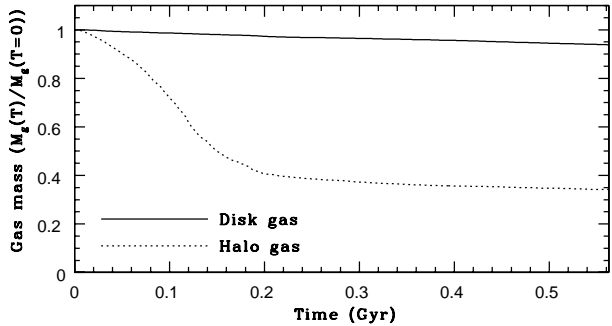,width=8.0cm}
\caption{
Time evolution of the total mass ($M_{\rm g}(T)$)
of disk (solid) and halo (dotted)
gas normalized to their initial values ($M_{\rm g} (T=0)$)
in the standard model.
}
\label{Figure. 3}
\end{figure}

\begin{figure}
\psfig{file=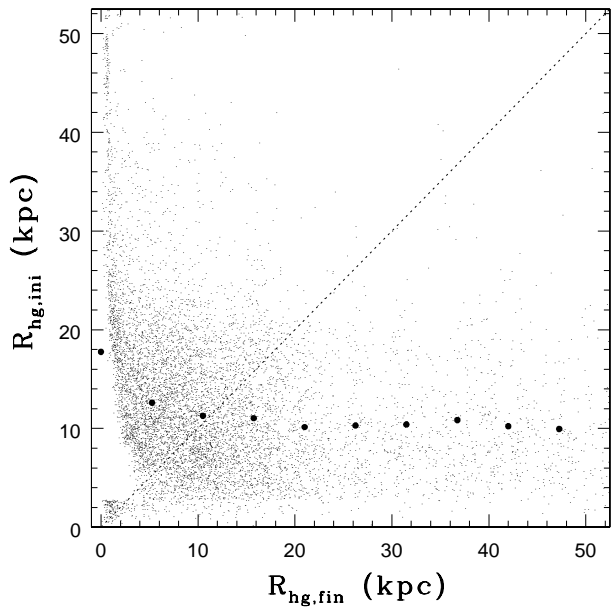,width=8.0cm}
\caption{
The initial distances ($R_{\rm hg,ini}$)
of halo SPH particles 
as a function of the final ones ($R_{\rm hg,fin}$)
at $T=0.56$ Gyr (i.e., final time step)
for the standard models.
A big filled circle represents the mean value of $R_{\rm hg,ini}$
for each $R_{\rm hg,fin}$ bin.
The dotted line shows $R_{\rm hg,fin}=R_{\rm hg,ini}$. 
Since SPH particles that can finally remain within the galaxy 
($R_{\rm hg,fin}<3R_{\rm d}$) are plotted in this figure,
the result means that the vast majority of
the remaining halo gas after ram pressure
stripping can originate from the inner halo ($R_{\rm hg,ini}\sim 10$ kpc).
This suggests that truncation of star formation after ram pressure
stripping of halo gas can proceed outside-in.
}
\label{Figure. 4}
\end{figure}

\begin{figure}
\psfig{file=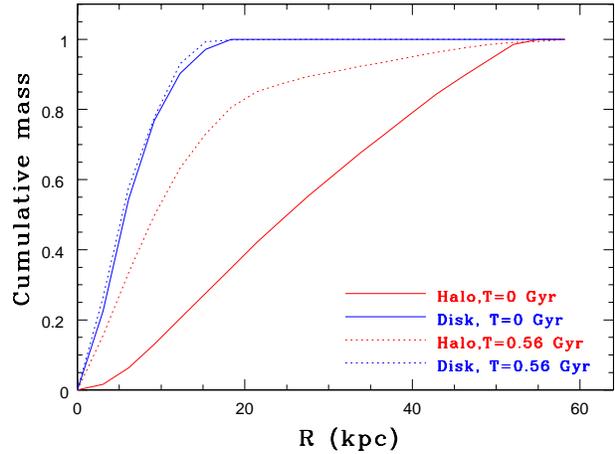,width=8.0cm}
\caption{
The cumulative mass distribution of halo (red) and disk (halo)
gas at $T=0$ Gyr (solid) and $T=0.56$ Gyr (dotted) in
the standard model.
The masses at $R$ are normalized to those at $R=3R_{\rm d}$.
}
\label{Figure. 5}
\end{figure}

\begin{figure}
\psfig{file=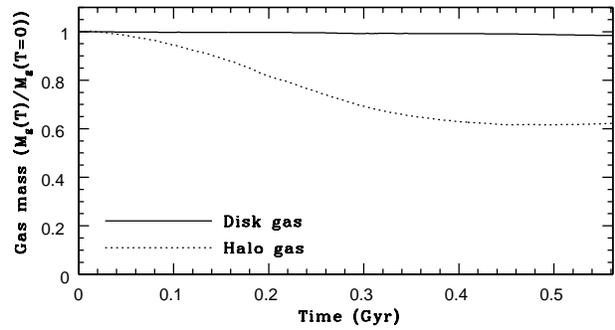,width=8.0cm}
\caption{
The same as Fig.3 but for the group model.
}
\label{Figure. 6}
\end{figure}

\begin{figure}
\psfig{file=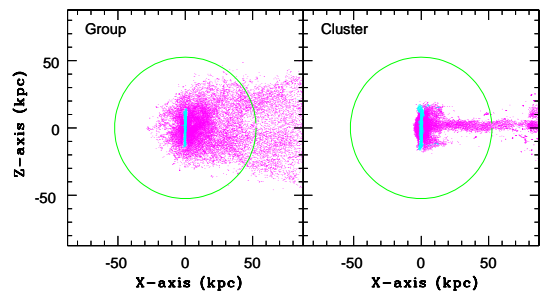,width=7.0cm}
\caption{
Final distributions of halo (magenta) and disk (cyan) gas
projected onto the $x$-$z$ plane for the group model
with $M_{\rm host}=10^{13} {\rm M}_{\odot}$ (left)
and the cluster one
with $M_{\rm host}=10^{14} {\rm M}_{\odot}$ (right).
The green solid circle in each frame represents the initial
size of the gaseous halo.
}
\label{Figure. 7}
\end{figure}

\section{Results}

\subsection{The standard model}

Fig. 2 shows how halo  gas  responses to  moderately strong
ram pressure of IGM in the standard model with
$M_{\rm host}=10^{14} {\rm M}_{\odot}$,
$T_{\rm IGM}=10^7$ K, and $V_{\rm r}=500$ km s$^{-1}$. 
The halo gas is efficiently stripped from
the disk galaxy to start to form a gaseous stream  behind the galaxy
within 0.14 Gyr. Owing to the presence of the disk gas,
halo gas initially located in the inner halo can be accumulated
above the disk and thus can not be stripped from the galaxy.
The gaseous stream with many small clumps formed after stripping
can finally become less remarkable within 0.56 Gyr and 
most of the remaining halo gas can be located close to the disk.
The final distribution of the halo gas appears to be 
more compact,  flattened, and inhomogeneous at $T=0.56$ Gyr.
The disk gas,  on the other hand, 
can not be stripped efficiently from the galaxy, 
because it is much more strongly
bounded by the disk in comparison with the halo gas.
Thus, ram pressure stripping of halo gas is much more effective
than that of disk gas.

Fig. 3 shows that 66\% and 6\% of initial  gas can be stripped
from halo and disk, respectively, by ram pressure within 0.56 Gyr
for the standard model. Fig. 4 shows that halo gas initially 
in the inner halo ($R_{\rm hg,ini}<10$ kpc)
is more likely to be still located in
the galaxy ($R_{\rm hg,fin}<R_{\rm halo}$). 
These remaining halo gas might well be accreted later onto the inner
region of the disk to increase slightly the total gas mass of the disk.
The halo gas initially located in the outer halo ($R>R_{\rm d}$) can be 
much more efficiently
stripped by ram pressure.
The stripping of the  outer halo gas 
might well severely suppress the accretion of the gas onto the outer
part of the galaxy and thus contribute to truncation of star formation
there.  

Fig. 4 shows that the radial distribution of the halo gas
can dramatically change after hydrodynamical interaction
between IGM and the halo gas, though
that of the disk gas
can hardly change owing to ram pressure weaker than the self-gravity
of the gas disk.
The distribution of halo gas becomes more compact
with the half-mass radius of $\sim 10$ kpc at $T=0.56$ Gyr,
because the stripping can happen much more efficiently
in the outer halo, where  the halo is more susceptible to
ram pressure effect owing to weaker gravity of the galaxy there.
These results in Figs. $2-5$ clearly demonstrate
that moderately  strong ram pressure of IGM 
can very efficiently remove the halo gas,
even though it can not remove the disk gas.

In order to understand more clearly how the disk gas in a disk
galaxy can 
promote or suppress the ram pressure stripping of the halo one,
we have investigated a comparative 
model with no disk gas (``no disk gas model'', $f_{\rm g}=0$).
It is found  that $F_{\rm strip}$
in the  no disk gas model is 0.79,
which is by a factor of 1.2 higher than that (0.66) in
the standard model with disk gas ($f_{\rm g}=0.1$).
The final distribution of the halo gas in the no disk gas model
is less flattened and a thin gaseous stream behind the  disk
can not be seen.
These results  mean that hydrodynamical interaction between disk and halo
gas can suppress ram pressure stripping of the halo gas.

As shown in Fig. 2,  the remaining halo gas at $T=0.56$ Gyr
is compressed and accumulated above the gas disk. About 28\% 
of the remaining halo gas can be located at $X<0$ (i.e., left side
of the gas disk) and is likely to be accreted finally onto the disk
for further star formation: the low-density gas located at
$X>0$ (i.e., right side of the disk) is less likely to be accreted
rapidly onto the disk.  Therefore, the mass fraction
of halo gas that can be used for later  star formation 
($F_{\rm sf}$) after ram pressure stripping 
can be significantly smaller than $1-F_{\rm strip}$ in disk
galaxies ($F_{\rm sf}$ can be
typically $\sim 0.3\times (1-F_{\rm strip}$)).

Interestingly, about 0.7\% of the halo gas
(corresponding roughly to $1.1 \times 10^7 {\rm M}_{\odot}$
within $R_{\rm halo}$)  can be fueled to
the central 1kpc of the disk.
This radial transfer of halo gas to the central region of 
the galaxy
can be seen in other models with different model parameters.
These  implies  that if the centrally accumulated  halo gas  can be further
transfered to the vicinity of the  massive black hole (MBH)
located in the center of the bulge, 
the halo gas can be used for fueling the MBH
with the accretion rate of $\sim 0.02 {\rm M}_{\odot}$ yr$^{-1}$.
These results imply that ram pressure of IGM can be responsible
for the activation of weak active galactic nuclei (AGN).

\begin{figure}
\psfig{file=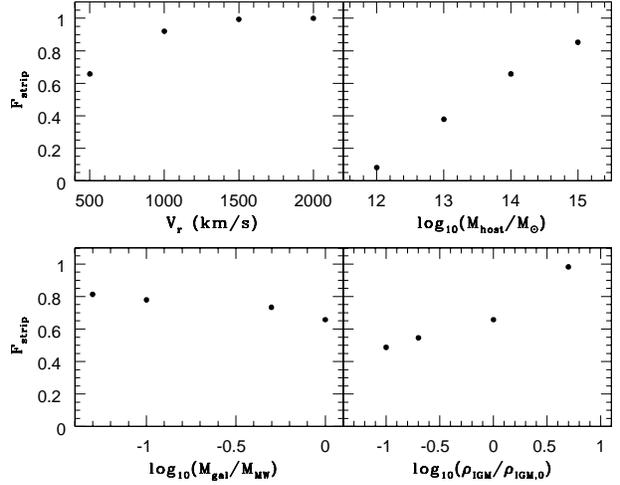,width=8.0cm}
\caption{
Parameter dependences of $F_{\rm strip}$ on
$V_{\rm r}$ (upper left),
$M_{\rm host}$ (upper right),
$M_{\rm gal}$ (lower left),
and ${\rho}_{\rm IGM}$ (lower right).
For each parameter dependence,
model parameters other than the described one
(e.g., $V_{\rm r}$ in the upper left frame)
are set to be the same as those used in the standard model.
In the lower right panel,
${\rho}_{\rm IGM, 0}$ is the same as ${\rho}_{\rm IGM}$
($=10^5 {\rm M}_{\odot}$ kpc$^{-3}$)
used in the standard model. 
}
\label{Figure. 8}
\end{figure}

\begin{figure}
\psfig{file=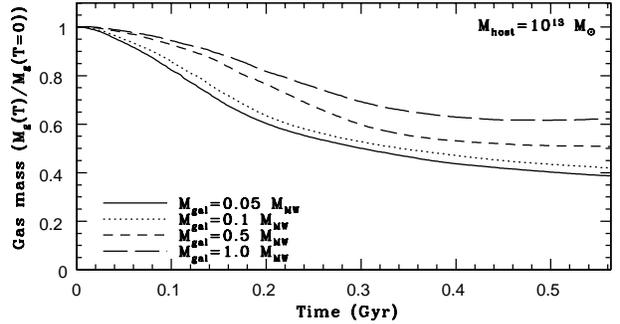,width=8.0cm}
\caption{
The same as Fig. 3 but for four different models
with different $M_{\rm gal}$ in the group
model with $M_{\rm host}=10^{13} {\rm M}_{\odot}$:
$M_{\rm gal}=0.05 M_{\rm MW}$ (solid),   
$M_{\rm gal}=0.1 M_{\rm MW}$ (dotted),
$M_{\rm gal}=0.5 M_{\rm MW}$ (short-dashed), 
and  $M_{\rm gal}=1.0 M_{\rm MW}$ (long-dashed). 
}
\label{Figure. 9}
\end{figure}

\begin{figure}
\psfig{file=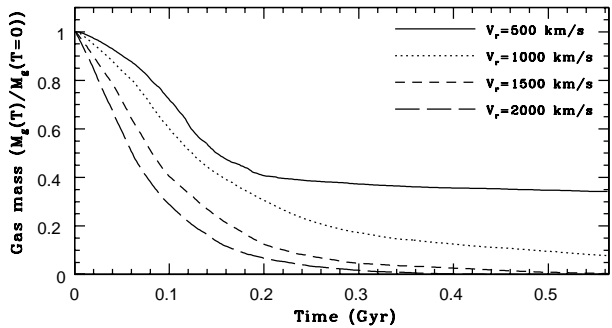,width=8.0cm}
\caption{
The same as Fig. 3 but for four different models
with different $V_{\rm r}$ in the cluster 
model with $M_{\rm host}=10^{14} {\rm M}_{\odot}$:
$V_{\rm r}=500$ km s$^{-1}$ (solid),   
$V_{\rm r}=1000$ km s$^{-1}$ (dotted),
$V_{\rm r}=1500$ km s$^{-1}$ (short-dashed), 
and $V_{\rm r}=2000$ km s$^{-1}$  (long-dashed). 
}
\label{Figure. 10}
\end{figure}

\subsection{Parameter dependences}

The dependences of $F_{\rm strip}$ and final radial distributions
of halo gas are described as follows:

(i)  Galactic halo gas can be efficiently stripped from disk galaxies
by ram pressure of IGM in groups with $M_{\rm host} \approx 10^{13}
{\rm M}_{\odot}$, 
though $F_{\rm strip}$ can be  smaller than 
those in cluster models with  $M_{\rm host} \approx 10^{14} {\rm
M}_{\odot}$. 
For example, as shown in Fig. 6,
the group model with $M_{\rm host}=10^{13} {\rm M}_{\odot}$
shows $F_{\rm strip}=0.38$, which is significantly smaller 
than that (0.66) in the cluster model with  
$M_{\rm host}=10^{14} {\rm M}_{\odot}$.
Fig. 7 clearly shows that the final distribution
of halo gas in the group model is less compact and less flattened than
that in the cluster model owing to weaker effects of ram pressure
of IGM in the group model.

(ii)  $F_{\rm strip}$ depends on $M_{\rm host}$ in such a way
that  $F_{\rm strip}$ is higher in models with higher $M_{\rm host}$.
As shown in Fig. 8,
the models with $M_{\rm host}=10^{12} {\rm M}_{\odot}$
(corresponding to small groups like the Local Group)
do not show any efficient ram pressure stripping of halo gas.
This result suggests that ram pressure stripping is not 
so important for galactic global star formation of 
luminous disk galaxies
in small groups with masses of $\sim 10^{12} {\rm M}_{\odot}$.

(iii) Less massive disk galaxies are likely to lose a larger
amount of their halo gas more rapidly for a given environment,
though the dependence of $F_{\rm strip}$ on $M_{\rm gal}$
is not very strong.
(See Fig. 8). 
Fig. 9 clearly shows that even in the group environment
with $M_{\rm host}=10^{13} {\rm M}_{\odot}$,
less massive galaxies can lose larger mass fractions of 
their halo gas more rapidly.
These results imply that less luminous disk galaxies
can truncate more rapidly their star formation after they enter into
group and cluster environments.

(iv) For a given $M_{\rm host}$,
$F_{\rm strip}$ are higher in  models with higher $V_{\rm r}$
owing to stronger force of ram pressure of IGM (See Fig. 8).
Also Fig.10 shoes that galactic halo gas can be more rapidly stripped from
disk galaxies by ram pressure in models with higher $V_{\rm r}$.
These results suggest that disk galaxies passing through
the inner regions of groups and clusters can truncate their star
formation more rapidly owing to more efficient  ram pressure stripping
of their halo gas.

(v) $F_{\rm strip}$  depends on ${\rho}_{\rm IGM}$ such that
$F_{\rm strip}$  is higher for models with higher   ${\rho}_{\rm
IGM}$ (See Fig. 8),
which is a natural result of $P_{\rm ram} \propto  {\rho}_{\rm IGM}
\times {V_{\rm r}}^2$. 
It is found that 
if ${\rho}_{\rm IGM} \sim 10^5 {\rm M}_{\odot}$ kpc$^{-3}$
and $V_{\rm r} \sim 400$ km s$^{-1}$
in the  small group model with $M_{\rm host}=10^{12} {\rm M}_{\odot}$
and $T_{\rm IGM}=10^6$ K,
then $F_{\rm strip}$ can be as high as 0.65. 
This result suggests that truncation of star formation
by ram pressure stripping of halo gas can be possible
even in small and/or compact groups,
if their IGM densities are as high as $10^5 {\rm M}_{\odot}$ kpc$^{-3}$
and member galaxies can have moderately larger velocity dispersion
($\sim 400$ km s$^{-1}$). 

(vi)  $F_{\rm strip}$  does not depend strongly on $f_{\rm g}$,
$\theta$, $\phi$, and
$R_{\rm halo}$. 
For example, $F_{\rm strip}$ can differ only by a factor of $\sim
1.2$ between models with $f_{\rm g}=0.01-0.1$.
Also  $F_{\rm strip}$ can differ  by a factor of $\sim
1.2$ between models with different $\theta$ and  $\phi$.
These results imply that initial distributions of halo and disk gas
are less important 
than other parameters (e.g., $V_{\rm r}$)
in determining $F_{\rm strip}$.

\section{Discussion}
\subsection{Comparison with previous works}

Bekki et al (2002) first investigated how much amount of galactic
halo gas can be stripped by ram pressure in groups and clusters of
galaxies based on rather idealized numerical simulations. 
They found that (i) about 90\% of galactic halo gas
can be stripped by the combination of tidal and ram pressure stripping
in clusters of galaxies, (ii) $F_{\rm strip}$ depends
on orbits of galaxies in groups and clusters,
and (iii) galactic halo gas can be effectively removed from
galaxies in groups only if the orbits of the galaxies are rather eccentric.
Although their results can be useful for better understanding
the origin of passive spirals, their models are not sophisticated
enough to discuss the dependences of  $F_{\rm strip}$
on galaxy environments in a quantitative manner.

Recently McCarthy et al. (2008) have conducted a thorough parameter
study on ram pressure stripping of hot gaseous halos in galaxies 
for groups and clusters of galaxies. Their numerical simulations
have shown that typically 70\% of the initial halo gas around galaxies
can be stripped by ram pressure within 10 Gyr in groups and clusters.
Although they did not include stellar disks and bulges in their
simulations, the result of $F_{\rm strip}$
is broadly  consistent with our simulations ($F_{\rm strip}\sim0.6-0.8$), 
in which
cosmologically motivated initial conditions of IGM and halo gas
are not adopted. The slight difference in $F_{\rm strip}$ between their
and our
simulations would be due simply to the fact that
they did not include disk gas, which is demonstrated to suppress
ram pressure stripping of galactic halo gas in the present study.
McCarthy et al. (2008) also have found that  $F_{\rm strip}$
is higher in galaxies with smaller masses for a cluster environment
with $M_{\rm host}=10^{14}$ $M_{\odot}$ (i.e., the models with
higher mass ratios show higher  $F_{\rm strip}$ in their Fig. 8).
This is also consistent qualitatively with our results.

Kawata \& Mulchaey (2008) have investigated how ram pressure of a groups
with $M_{\rm host}=8\times 10^{12} {\rm M}_{\odot}$ influences
gaseous components in disk galaxies with maximum circular velocities
of $\sim 150$ km s$^{-1}$ based on cosmological
chemodynamical numerical simulations.  They have found that ram pressure in
the group can be more than enough to remove hot halo gas of the
galaxies,
though it can not remove their cold gas within disks.
These results are consistent qualitatively with the present ones,
though initial conditions of disk galaxies in groups are different
between Kawata \& Mulchaey (2008)  and the present study. 
This consistency implies that ram pressure of IGM in groups 
with $M_{\rm host} \sim 10^{13} {\rm M}_{\odot}$ 
can play an important role in controlling gas accretion from halos
onto galactic disks and thus determining star formation histories within
them.

Although broadly consistent results between the present and previous
results clearly demonstrate 
that evolution of hot halo gas around galaxies under
ram pressure of IGM can be one of key determinants for galaxy evolution,
it is not so  unclear how global properties
of galaxies, in particular,  their morphological,
structural,  and kinematical properties,
change after removal of their halo gas.  
The present and previous simulations do not enable us to
investigate the details of dynamical properties of disk galaxies
after ram pressure stripping
in a fully self-consistent manner.
Therefore it is still unclear whether the observed rapid evolution
of S0 fraction in groups (e.g.,  Wilman et al. 2009) can result from  
truncation of star formation in spirals owing to the removal
of halo gas in group environments. 
Given that minor and unequal-mass merging can create S0s (e.g., Bekki
1998),  it is important for our future studies to confirm whether 
the simulated
physical properties of S0s formed from spirals through removal of
their halo gas can reproduce the observed ones reasonably well.

\subsection{Implications of the present results}

\subsubsection{Outside-in truncation of star formation}

The present study has demonstrated that outer halo gas can be
preferentially stripped from disk galaxies during hydrodynamical
interaction between hot IGM and halo gas: the remaining gas around
disks mostly originates from inner halos.  This result implies that
if halo gas of a galaxy has intrinsic angular momentum to be accreted
onto the disk,   the preferential removal of the outer gas 
might well result in the suppression
of the growth of the outer disk of the galaxy: the inner disk would
grow after ram pressure stripping, though the accretion rate of halo
gas would be significantly reduced.
Our previous work (Bekki et al. 2002) showed that after removal
of halo gas,  stellar velocity dispersions (thus the $Q$ parameter)
of a disk galaxy can significantly increase in its outer part
owing to dynamical
heating of the disk by spiral arms.
The present result combined with our previous one
therefore suggests that gradual truncation of star formation
can proceed outside-in
in disk galaxies
owing to more significantly increased $Q$ and reduced $f_{\rm g}$
(i.e., gas mass fraction)  in 
their outer disks.

This possible outside-in truncation scenario of star formation  can 
provide a clue to the origin of dusty star-forming regions observed
in passive spiral galaxies in distant groups and clusters (e.g., Dressler
et al. 2009).
Recent Spitzer observations of passive spiral galaxies
have shown that passive spirals, 
 which were previously suggested
to be ``dead and red'' disk galaxies (Couch et al. 1998), 
show some levels of star
formation  (Dressler et al. 2009).
  The observed dusty star-forming, optically red spirals
mean that star formation activities can be heavily obscured by dust.
It is not clear why passive spirals, which should be gas-poor galaxies
to explain the observed red colors (i.e., low-level star formation),
can show heavily observed star-forming regions  (which requires
high-density gas and dust). 
The present outside-in truncation implies that star formation
can continue only in the inner high-density regions (after  
stripping of halo gas), where dust extinction is highly likely to be
significant. Thus the outside-in truncation scenario can naturally
explain why passive spirals can have dusty-star forming regions
without showing star-forming regions not obscured by dust in their
outer parts.

Typically $\sim 5$\% of halo gas can be accumulated above gas disks
during ram pressure stripping and these gas can be strongly
compressed and thus strongly interact hydrodynamically with the gas disks
(see the Appendix A for the details for the evolution of
gas disks during ram pressure stripping). 
As shown in previous numerical simulations
(e.g., Bekki \& Couch 2003; Kronberger et al. 2008),
external high-pressure of gas may well trigger efficient star formation
in gas disks.
Therefore, the present results imply that before gradual truncation of star
formation due to  ram pressure stripping of halo gas in disk galaxies,
star formation in the galaxies
could increase to some extent.
Our future more sophisticated simulations including star formation
within giant molecular clouds in disk galaxies under ram pressure of IGM
will enable us to discuss this problem in a more quantitative way.

\subsubsection{Recycling of ISM}

Recent numerical simulations have shown that 
metal-rich gas ejected from massive OB stars and supernovae
formed during efficient star formation in disk galaxies
can be accreted onto the thin disks owing to hydrodynamical
interaction between the gaseous ejecta and the gaseous halos
(Bekki et al. 2009).
This physical mechanism  of galactic halo gas
to keep gaseous ejecta within galaxies 
can effectively work only if the densities of gaseous halos
can be as high as $10^{-5}$ ${\rm cm}^{-3}$ (Bekki et al. 2009).
These results combined with the present ones therefore suggest that
if galaxies enter inter group and cluster environments and lose most
of their halo gas owing to ram pressure stripping of the halo gas,
then recycling processes of metal-rich gaseous
ejecta in interstellar medium (ISM) of galaxies 
can be dramatically changed owing to gaseous halos with much
lower densities.

If recycling processes of metal-enriched ISM can be severely suppressed
by ram pressure stripping of galactic halo gas,
then chemical evolution of disk galaxies after the stripping
can be also significantly
changed.  For example, energetic stellar winds from supernovae
and massive OB stars
in luminous disk galaxies under the influence of ram pressure
can easily escape from the galaxies and be dispersed into
IGM of their host environments.
Less energetic winds like those from AGB stars, on the other hand, can be still
trapped within the galaxies so that their ejecta can be used for
further star formation and thus for chemical evolution. 
As a result of these,  abundance patterns might well be significantly
different between field disk galaxies, which can retain gaseous ejecta
both from AGB stars and supernovae well, and cluster/group disk
galaxies, which can retain only AGB ejecta.
Our future 
quantitative investigation on chemical evolution of disk galaxies
after ram pressure stripping of their halo gas
will enable us to address this important question as to how abundance
patterns in disks are  different between field and group/cluster
disk galaxies.

\section{Conclusions}

We have investigated time evolution of galactic halo gas
in  disk galaxies under moderately strong ram pressure of IGM in 
different environments based on self-consistent hydrodynamical
simulations with variously different model parameters. We summarize our
principle results as
follows.

(1) Even moderately strong ram pressure of IGM with ${\rho}_{IGM}
\approx 10^5$ ${\rm M}_{\odot}$  kpc$^{-3}$ and $V_{\rm r}\approx 500$
km s$^{-1}$ can strip galactic halo gas efficiently in clusters of
galaxies with $M_{\rm host}=10^{14}$ $M_{\odot}$.
Typically $60-80$\% of initial halo gas can be stripped from disk
galaxies (i.e., $F_{\rm strip}=0.6=0.8$) in clusters,
which is  broadly consistent with results
of other studies. 

(2)  $F_{\rm strip}$ depends on $M_{\rm host}$ such that
it can be higher in  environments with higher  $M_{\rm host}$.
For example,  $F_{\rm strip}$ can be as large as 0.4 in group
environments
with $M_{\rm host}=10^{13}$ $M_{\odot}$.  
$F_{\rm strip}$ in small and compact groups
with low IGM temperature ($T_{\rm IGM}=10^6$ K)  can be as large as $\sim
0.6$, if the groups have IGM with
${\rho}_{IGM}\approx 10^5$ ${\rm M}_{\odot}$  kpc$^{-3}$ and
$V_{\rm r}\approx 400$ km s$^{-1}$.
These results imply that strangulation can happen even in 
small and/or
compact groups with higher velocity dispersions.

(3) The presence of disk gas in disk galaxies
can suppress the ram pressure stripping
of their halo gas owing to hydrodynamical interaction between halo and disk
gas. This result suggests that stripping processes of halo gas by ram
pressure can be significantly different between late- and early-type
galaxies with and without disk gas.

(4) Disk galaxies with lower masses ($M_{\rm gal}$)
are likely to show appreciably
higher  $F_{\rm strip}$ in a given environment. This result implies
that strangulation can proceed more rapidly and more efficiently
in disk galaxies with lower $M_{\rm gal}$.
The remaining halo gas of disk galaxies after
ram pressure stripping shows  more compact and flattened distributions
and has
jellyfish-like configurations, irrespective of $M_{\rm gal}$.

(5) Most halo gas stripped by ram pressure in disk galaxies can originate from
the outer parts of the halos so that only halo gas initially in the inner
parts can be accreted onto the disks during/after ram pressure stripping.
This suggests that strangulation is more efficient in outer disks and
thus that truncation of star formation by stripping
of halo gas may well proceed outside-in in disk galaxies.
This outside-in truncation can provide a clue to the origin of 
red passive spirals in distant groups and clusters.

(6) Ram pressure stripping of galactic halo gas is much more efficient
than that of disk gas in  groups and clusters and efficiencies of
halo and disk gas stripping can be different in different environments
with different $M_{\rm host}$, $V_{\rm r}$, and  ${\rho}_{\rm IGM}$.
Thus the mass ratios of halo gas to disk one in disk galaxies can
reflect  in what environments they have resided in their histories 
and thus be significantly different between disk galaxies.

(7) It is suggested that recycling processes of ISM
and chemical evolution of galaxies 
significantly change after removal of galactic halo gas.
One example is that gaseous ejecta from energetic massive OB stars
and supernovae can more easily  escape from disk galaxies
after halo gas stripping
owing to much less effective hydrodynamical interaction between
gaseous  ejecta and halo gas.  Abundance patterns 
of ISM in disk galaxies are thus suggested to be significantly 
different between different environments.

(8) Small fractions of halo gas ($0.3-1$\%) in disk galaxies can be
accumulated into 
the central 1 kpc of galactic bulges owing to the compression of
halo gas during ram pressure stripping.
These gas can be  used for fueling the central MBH and activating
weak AGN there,
though the final dynamical fate of the accumulated halo  gas is beyond the scope
of this paper.

\section{Acknowledgment}
I am   grateful to the anonymous referee for valuable comments,
which contribute to improve the present paper.
KB acknowledges the financial support of the Australian Research
Council
throughout the course of this work.
I am   grateful to participants in the last conference
``galaxy evolution and environment'' held in Kuala Lumpure on 30 March
$-$ 3 April 2009 for their having extensive discussion on galaxy evolution
in groups and clusters with me,
which helped  me better interpret the present numerical results
in terms of the corresponding observational ones.
Numerical computations reported here were carried out on the GRAPE
system
at the University of New South Wales  and that
kindly made available by the Center for computational
astrophysics
(CfCA) of the National Astronomical Observatory of Japan.

\appendix

\begin{figure}
\psfig{file=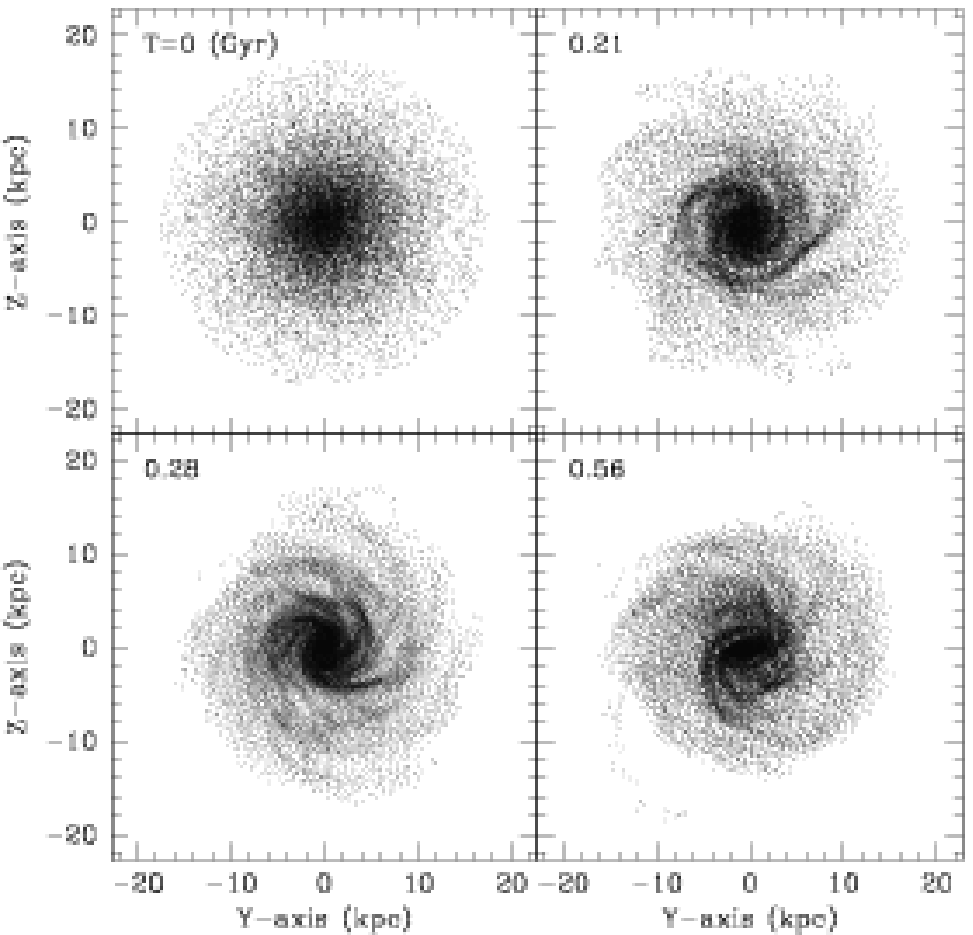,width=8.0cm}
\caption{
Morphological evolution of disk  gas
projected onto the $y$-$z$ plan
in a disk galaxy for the standard model. 
The time $T$ in units of Gyr is shown in the upper left corner
for each frame.
}
\label{Figure. 11}
\end{figure}

\begin{figure}
\psfig{file=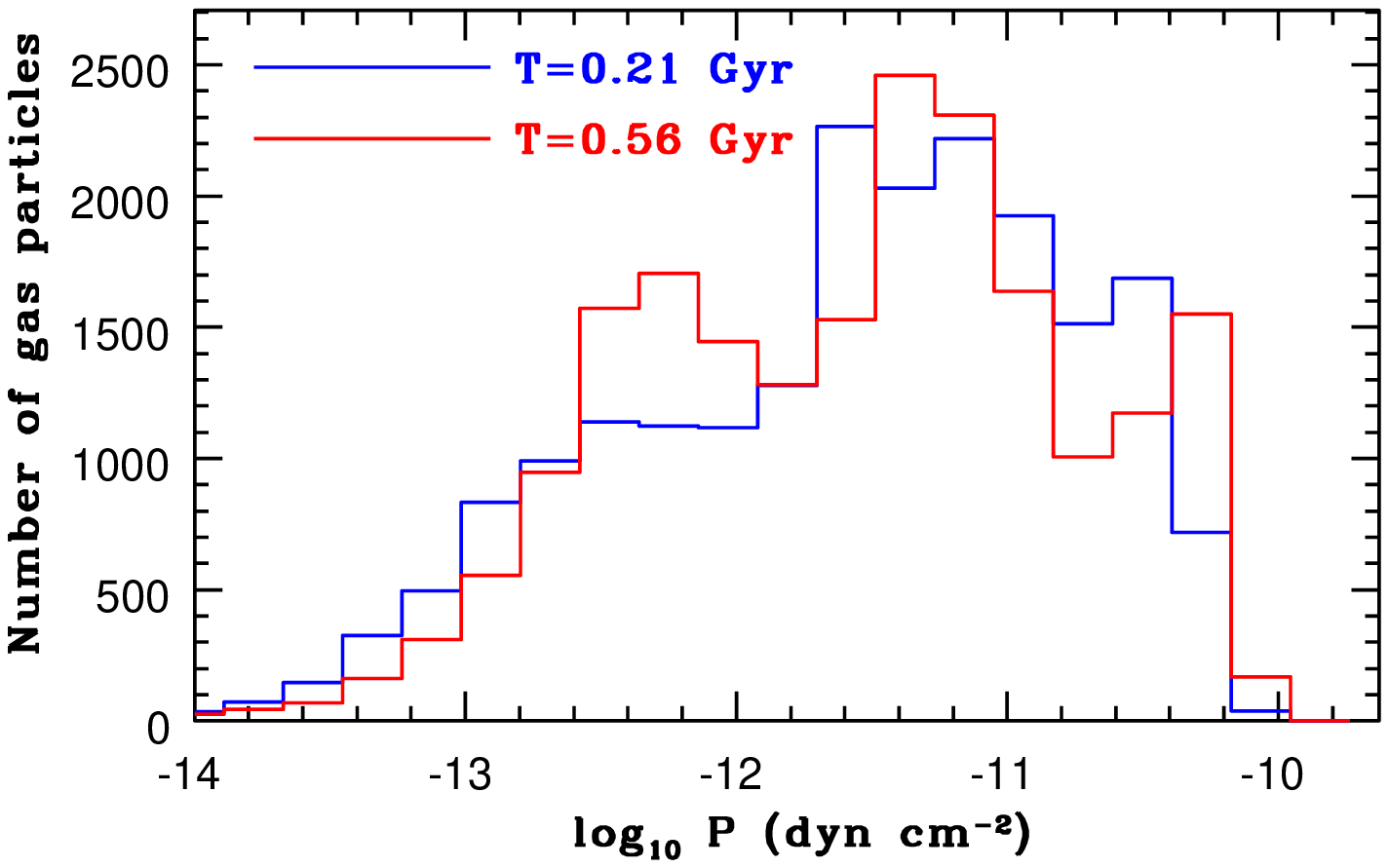,width=8.0cm}
\caption{
The number distributions of gaseous  pressure ($P$)  for
disk gas particles at $T=0.21$ Gyr (blue) and $T=0.56$ Gyr (red)
in the standard model. 
}
\label{Figure. 12}
\end{figure}

\section{Physical properties of gas disks during ram pressure stripping}

Fig. A1 shows the time evolution of morphological properties
of the gas disk for the last 0.56 Gyr in the standard model. 
As the halo gas is compressed by the strong ram pressure of IGM,
hydrodynamical interaction between disk and halo gas becomes 
stronger ($T=0.21$ Gyr). Gaseous spiral arms can be developed
in the inner region of the disk during ram pressure stripping,
though the outer part of the gas disk
can lose a minor fraction of its initial gas ($T=0.28$ Gyr). 
The final morphology of the gas disk appears to be more compact 
($T=0.56$ Gyr) owing to compression of the gas disk
by the halo gas and IGM.

Fig. A2 shows the number distributions of pressure of disk gas particles
($P$) at $T=0.21$ Gyr and $T=0.56$ Gyr in the standard model.
The mean $P$ at $T=0.21$ Gyr and $T=0.56$ Gyr are
$1.1 \times 10^{-11}$ dyn cm$^{-2}$ and $1.3 \times 10^{-11}$ dyn
cm$^{-2}$, respectively.
(i.e., mean ${\log}_{10} P$ 
at $T=0.21$ Gyr and $T=0.56$ Gyr are 
$-10.94$ and $-10.88$, respectively).
Although the mean pressure is not so different between
the two time steps,
the mass fraction of disk gas particles that have pressure
higher than $5 \times 10^{-11}$ dyn cm$^{-2}$ ($f_{\rm thres}$) 
is significantly different: $f_{\rm thres}$=0.04 at $T=0.21$ Gyr
and 0.09 at $T=0.56$ Gyr.
This would reflect the fact that as time passes by,
the disk gas particles can feel stronger pressure of
halo gas (and IGM) owing to the more strongly  compressed halo gas.

Recently Kapferer et al. (2009) have shown that if  the pressure
of IGM is as high as $5 \times 10^{-11}$ dyn cm$^{-2}$,
ram pressure can enhance significantly star formation in disk
galaxies owing to compression of disk gas by ram pressure force.
Our results in Fig. A2 therefore imply that star formation 
can be enhanced for some minor fraction ($<10$\%) of disk gas
during ram pressure stripping.
Although numerical models 
$P$ in  Kapfere et al. (2009) and in the preset
study are different, 
our results in Fig. A2 confirm the early suggestion
(e.g., Bekki \& Couch 2003; Kronberger et al. 2008)
that ram pressure can trigger formation of new stars in gas disks.
Numerical results by Kronberger et al. (2008)
and by the present study are consistent with each other
in that they show more efficient stripping of disk gas
in outer parts of gas disks:
star formation might well be truncated outside-in.

\section{Description of the animation}

We have made a simple animation for the time evolution of
mass distributions of halo and disk gas projected onto the $x$-$z$ plane 
for the standard model.
Only halo and gas particles
are shown by magenta and cyan,
respectively for clarity: Dark matter halo, stellar disk and bulge,
and hot IGM are not shown. The model parameters for the standard model
are described in detail in the main text.

Hot gas particles for IGM are initially placed on equally spaced grids
so that the adopted uniform density distribution of IGM can be achieved
in the present study.
In the early phase of ram pressure stripping of galactic halo gas by IGM,
some grid-like local structures can be seen owing to (i) the adopted
IGM distribution and (ii) smaller number of the IGM particle. 
Since these local structures can soon disappear,  
the final mass and distributions of galactic halo gas after ram pressure 
stripping, which are the main subjects of the present paper,
can not be influenced  significantly by the development of such structures
(which is due to the adopted numerical models for IGM).
The movie is available with the online version of the
paper.

\end{document}